\documentclass[twocolumn,preprintnumbers,superscriptaddress,prl,nofootinbib]{revtex4}
\usepackage{graphicx}
\usepackage{times}
\usepackage{epsfig}
\usepackage{amsmath}
\usepackage{amsfonts}
\usepackage{amssymb}
\usepackage{url}
\usepackage{subfigure}
\newcommand{\be}{\begin{equation}}
\newcommand{\ee}{\end{equation}}
\newcommand{\bea}{\begin{eqnarray}}
\newcommand{\eea}{\end{eqnarray}}

\begin{document}
\preprint{OU-HEP-994}
% Page numbers bottom-center
\pagestyle{plain}
%%%%%%%%%%%%%%%%%%%%%%%%%%%%%%%%%%%%%%%%%%%%%%%%%%%%%%%%%%%%
\title{
Probing the seesaw mechanism at the 250 GeV ILC
}
%%%%%%%%%%%%%%%%%%%%%%%%%%%%%%%%%%%%%%%%%%%%%%%%%%%%%%%%%%%%%
\author{
Arindam Das
}
%\email{arindam.das@het.phys.sci.osaka-u.ac.jp},
\affiliation{
Department of Physics, 
Osaka University, Toyonaka, 
Osaka 560-0043, Japan
}
\author{Nobuchika Okada}
%\email{okadan@ua.edu}
\affiliation{
Department of Physics and Astronomy,
University of Alabama,
Tuscaloosa, AL 35487, USA
}
\author{Satomi Okada}
%\email{satomi.okada@ua.edu}
\affiliation{
Department of Physics and Astronomy,
University of Alabama,
Tuscaloosa, AL 35487, USA
}
\author{Digesh Raut}
%\Email{draut@udel.edu}
\affiliation{
Bartol Research Institute,
Department of Physics and Astronomy,
University of Delaware, Newark, DE 19716, USA
}
%\date{\today}
%\baselineskip 36pt

%%%%%%%%%%%%%%%%%%%%%%%%%%%%%%%%%%%%%%%%%%%%%%%%%%%%%%%%%%%%
\begin{abstract}

We consider a gauged U(1)$_{B-L}$ (Baryon-minus-Lepton number) extension of the Standard Model (SM), 
  which is anomaly-free in the presence of three Right-Handed Neutrinos (RHNs). 
Associated with the U(1)$_{B-L}$ symmetry breaking  
  the RHNs acquire their Majorana masses and then play the crucial role to generate the neutrino mass matrix 
  by the seesaw mechanism.   
Towards the experimental confirmation of the seesaw mechanism, 
  we investigate a RHN pair production through the U(1)$_{B-L}$ gauge boson ($Z^\prime$) 
  at the 250 GeV International Linear Collider (ILC).  
The $Z^\prime$ gauge boson has been searched at the Large Hadron Collider (LHC) Run-2 
   and its production cross section is already severely constrained. 
The constraint will become more stringent by the future experiments 
   with the High-Luminosity upgrade of the LHC (HL-LHC). 
We find a possibility that even after a null $Z^\prime$ boson search result at the HL-LHC, 
   the 250 GeV ILC can search for the RHN pair production 
   through the final state with same-sign dileptons plus jets, 
   which is a ``smoking-gun'' signature from the Majorana nature of RHNs. 
In addition, some of RHNs are long-lived and leave a clean signature with a displaced vertex. 
Therefore, the 250 GeV ILC can operate as not only a Higgs Factory
   but also a RHN discovery machine to explore the origin of the Majorana neutrino 
   mass generation, namely the seesaw mechanism.

\end{abstract}
\maketitle

%%%%%%%%%%%%%%%%%%%%%%%%
% Main Boday
%%%%%%%%%%%%%%%%%%%%%%%%

Type-I seesaw \cite{seesaw} is probably the simplest mechanism 
   to naturally generate tiny masses for the neutrinos in the Standard Model (SM),  
   where Right-Handed Neutrinos (RHNs) with large Majorana masses play the crucial role. 
It has been known for a long time that the RHNs are naturally incorporated into 
   the so-called minimal $B-L$ model \cite{MBL}, 
   in which the global U(1)$_{B-L}$ (Baryon-minus-Lepton number) symmetry in the SM is gauged. 
In addition to the SM particle content, the model contains a minimal new particle content, 
   namely, the U(1)$_{B-L}$ gauge boson ($Z^\prime$ boson), 
   three RHNs, %with a $B-L$ charge $-1$, 
   and a U(1)$_{B-L}$ Higgs field. % with a $B-L$ charge $+2$. 
The existence of the three RHNs is crucial for the model to be free from all the gauge and 
   the mixed gauge-gravitational anomalies. 
Associated with the U(1)$_{B-L}$ symmetry breaking triggered by the vacuum expectation value (VEV) 
   of the $B-L$ Higgs field, 
   the $Z^\prime$ boson mass and the Majorana masses for the RHNs are generated. 
Once the electroweak symmetry is broken, the type-I seesaw mechanism generates 
    the mass matrix for the light SM neutrinos.

If the seesaw scale, in other words, the mass scale of the Majorana RHNs lies at the TeV scale or lower, 
   RHNs (more precisely, heavy Majorana neutrino mass eigenstates after the seesaw mechanism) 
   can be produced at high energy colliders through the process mediated by the $Z^\prime$ boson. 
Once produced, the RHN decays into the SM particles through the SM weak gauge boson 
   or the SM Higgs boson mediated processes. 
Since the RHNs are originally singlet under the SM gauge group, 
   the decay process implies a mass mixing between the RHNs and the SM neutrinos 
   through the type-I seesaw mechanism. 
Among possible final states from the RHN decay, 
   it is of particular interest to consider the same-sign dilepton final state,   
   a ``smoking-gun'' signature for the RHNs Majorana nature, for which the SM backgrounds are few. 
For prospects of discovering the Majorana RHN at the future Large Hadron Collider (LHC) experiments, 
   see, for example, Refs.~\cite{Kang:2015uoc, Cox:2017eme, Accomando:2017qcs, Das:2017flq, Das:2017deo, Jana:2018rdf}.

In this paper, we investigate the RHN production at future $e^+ e^-$ colliders, in particular, 
   the International Linear Collider (ILC), in the context of  a gauged $B-L$ extension of the SM. 
The ILC is proposed with a staged machine design, with the first stage at 250 GeV 
   with a luminosity goal of 2000/fb \cite{Fujii:2017vwa}. 
Setting the ILC energy at 250 GeV maximizes the SM Higgs boson production cross section, 
   and hence the 250 GeV ILC will be operating as a Higgs Factory. x
This machine allows us to precisely measure the Higgs boson properties to test the SM Higgs sector. 
One may think that the 250 GeV ILC is not a powerful machine compared to the LHC 
   in exploring new physics if it is less related to the SM Higgs sector. 
Given the present status with no evidence of new physics in the LHC data 
   and the prospect of new physics search at the High-Luminosity LHC (HL-LHC) in the near future, 
   it seems quite non-trivial to consider new physics for which the 250 GeV ILC is more capable than the HL-LHC. 
The main point of this paper is to show that the 250 GeV ILC can, in fact, probe the RHN pair production 
   mediated by the $Z^\prime$ boson even in the worst case scenario 
   that the HL-LHC data with a goal 3000/fb luminosity would show no evidence 
   for a resonant $Z^\prime$ boson production.

%%%%%%%%%%%%%%%%%%%%%%%%%%%%%%%%%%%%%%%%%%%%%%%%%%%
\begin{table}[t]
\begin{center}
\begin{tabular}{c|ccc|c|c}
      &  SU(3)$_C$  & SU(2)$_L$ & U(1)$_Y$ & U(1)$_{B-L}$ \\ 
\hline
$q^{i}_{L}$ & {\bf 3 }    &  {\bf 2}         & $ 1/6$       &  $1/3  $ \\
$u^{i}_{R}$ & {\bf 3 }    &  {\bf 1}         & $ 2/3$       & $1/3  $ \\
$d^{i}_{R}$ & {\bf 3 }    &  {\bf 1}         & $-1/3$       & $1/3 $\\
\hline
$\ell^{i}_{L}$ & {\bf 1 }    &  {\bf 2}         & $-1/2$       & $-1   $ \\
$e^{i}_{R}$    & {\bf 1 }    &  {\bf 1}         & $-1$         & $-1 $ \\
\hline
$H$            & {\bf 1 }    &  {\bf 2}         & $- 1/2$       & $0$ \\  
\hline
\hline
$N^{i}_{R}$    & {\bf 1 }    &  {\bf 1}         &$0$                    & $- 1 $ \\ 
\hline
$\Phi        $    & {\bf 1 }    &  {\bf 1}         &$0$                    & $+2 $ \\ 
\end{tabular}
\end{center}
\caption{
The particle content of the minimal $B-L$ model. 
In addition to the three generations of SM particles ($i=1,2,3$), 
  three RHNs ($N^i_R$) and one $B-L$ Higgs field ($\Phi$) are introduced.  
}
\label{table1}
\end{table}
%%%%%%%%%%%%%%%%%%%%%%%%%%%%%%%%%%%%%%%%%%%%%

In this paper, we consider two simple gauged $B-L$ extended SMs. 
One is the minimal $B-L$ model whose particle content is listed in Table \ref{table1}.  
%In addition to the three generations of the SM quarks and leptons, 
%   three RHNs and one $B-L$ Higgs field are introduced. 
The model is free from all the gauge and the mixed gauge-gravitational anomalies,  
   thanks to the presence of the three RHNs. 
   
In addition to the SM, we introduce Yukawa couplings involving new fields: 
\bea
\mathcal{L} _{Y} = 
-\sum_{i,j=1}^{3} Y_{D}^{ij} \overline{\ell_{L}^{i}} H N_{R}^{j}
-\frac{1}{2} \sum_{k=1}^{3} Y_{N}^{k} \Phi \overline{N_{R}^{k \,c}} N_{R}^{k} + \rm{h. c.}, 
\label{Yukawa}
\eea
where three RHNs ($N_R^j$) have the Dirac Yukawa couplings with the SM lepton doublets 
   as well as  the Majorana Yukawa couplings with the $B-L$ Higgs field.  
We assume a suitable Higgs potential to yield VEVs for the Higgs fields, 
  $\langle H \rangle = (v/\sqrt{2}, 0)^T$ with $v=246$ GeV and $\langle \Phi \rangle = v_\phi/\sqrt{2}$, 
  to break the electroweak and the U(1)$_{B-L}$ symmetries, respectively. 
The symmetry breakings generate the $Z^\prime$ boson mass, 
  the Majorana masses for RHNs, and the neutrino Dirac masses: 
\bea
 m_{Z^\prime} = 2 \, g_{BL} \, v_\phi, \; \;
 m_{N^j} &=& \frac{Y_N^{j}}{\sqrt{2}} v_\phi,  \; \;  
 m_{D}^{ij}=\frac{Y_{D}^{ij}}{\sqrt{2}} v,  
\label{mass_minimal} 
\eea
where $g_{BL}$ is the $B-L$ gauge coupling.

%%%%%%%%%%%%%%%%%%%%%%%%%%%%%%%%%%%%%%%%%%%%%%%%%%%
\begin{table}[t]
\begin{center}
\begin{tabular}{c|ccc|c|c}
      &  SU(3)$_C$  & SU(2)$_L$ & U(1)$_Y$ & U(1)$_{B-L}$ \\ 
\hline
%$q^{i}_{L}$ & {\bf 3 }    &  {\bf 2}         & $ 1/6$       &  $1/3  $ \\
%$u^{i}_{R}$ & {\bf 3 }    &  {\bf 1}         & $ 2/3$       & $1/3  $ \\
%$d^{i}_{R}$ & {\bf 3 }    &  {\bf 1}         & $-1/3$       & $1/3 $\\
%\hline
%$\ell^{i}_{L}$ & {\bf 1 }    &  {\bf 2}         & $-1/2$       & $-1   $ \\
%$e^{i}_{R}$    & {\bf 1 }    &  {\bf 1}         & $-1$         & $-1 $ \\
%\hline
%$H$            & {\bf 1 }    &  {\bf 2}         & $- 1/2$       & $0$ \\  
%\hline
%\hline
$N^{1, 2}_{R}$    & {\bf 1 }    &  {\bf 1}         &$0$                    & $- 4 $ \\ 
$N_R^3$    & {\bf 1 }    &  {\bf 1}         &$0$                           & $+ 5 $   \\
\hline
$H_\nu$            & {\bf 1 }       &  {\bf 2}       &$ -\frac{1}{2}$                  & $ 3 $  \\ 
$\Phi_A$            & {\bf 1 }       &  {\bf 1}       &$ 0$                  & $ +8  $  \\ 
$\Phi_B$            & {\bf 1 }       &  {\bf 1}       &$ 0$                  & $ -10 $  \\ 
\end{tabular}
\end{center}
\caption{
New particle content of the alternative $B-L$ model. 
The three RHNs and the $B-L$ Higgs filed in Table \ref{table1} 
are replaced by three RHNs ($N_R^{1,2,3}$) with flavor-dependent charges 
  and three Higgs fields ($H_\nu, \Phi_{A,B}$).   
}
\label{table2}
\end{table}
%%%%%%%%%%%%%%%%%%%%%%%%%%%%%%%%%%%%%%%%%%%%%

The other model is what we call ``alternative $B-L$ model'', 
   in which a  U(1)$_{B-L}$ charge $-4$ is assigned to two RHNs ($N_R^{1,2}$),   
   while a U(1)$_{B-L}$ charge $-5$ is assigned for the third RHN ($N_R^3$) \cite{AltU1X}.
The cancellation of all the gauge and the mixed gauge-gravitational anomalies is achieved 
   also by this charge assignment. 
In the alternative $B-L$ model, we introduce a minimal Higgs sector 
   with one new Higgs doublet $H_\nu$ and two U(1)$_{B-L}$ Higgs fields $\Phi_{A,B}$. 
Table~\ref{table2} lists the new particle content.

In the alternative $B-L$ model, we introduce the following Yukawa couplings involving new fields: 
\bea
\mathcal{L} _{Y} &=& 
-\sum_{i=1}^{3} \sum_{j=1}^{2} Y_{D}^{ij} \overline{\ell_{L}^{i}} H_\nu N_{R}^{j} 
-\frac{1}{2} \sum_{k=1}^{2} Y_{N}^{k} \Phi_A \overline{N_{R}^{k \,c}} N_{R}^{k} \nonumber \\
&&-\frac{1}{2} Y_{N}^{3} \Phi_B \overline{N_{R}^{3 \,c}} N_{R}^{3}
+ \rm{h. c.}.  
\label{ExoticYukawa}
\eea
%where $N_R^3$ has only the Majorana Yukawa coupling because of the gauge invariance. 
We assume a suitable Higgs potential to trigger the gauge symmetry breaking 
  with non-zero VEVs as 
  $\langle H \rangle = (v_h/\sqrt{2},  0)^T$, 
  $\langle H_\nu \rangle = (v_\nu/\sqrt{2},  0)^T$, 
  and $\langle \Phi_{A, B} \rangle =  v_{A, B}/\sqrt{2}$, 
where we require $\sqrt{v_h^2 + v_\nu^2} = 246$ GeV for the electroweak symmetry breaking. 
After the U(1)$_{B-L}$ and SM gauge symmetries are spontaneously broken, 
  the $Z^\prime$ boson mass, the Majorana masses for the RHNs, 
  and the Dirac neutrino masses are generated:  
\bea
m_{Z^\prime} &=& g_{BL} \sqrt{64 v_{A}^2+ 100 v_{B}^2 + 9 v_\nu^2 } \simeq  g_{BL} \sqrt{64 v_{A}^2+ 100 v_{B}^2}, \nonumber \\
m_{N^{1,2}}&=&\frac{Y_N^{1,2}}{\sqrt{2}} v_A,  \; \; 
m_{N^3}=\frac{Y_N^3}{\sqrt{2}} v_B,  \; \; 
m_{D}^{ij}=\frac{Y_{D}^{ij}}{\sqrt{2}} v_\nu.  
\label{mass_alternative}
\eea 
Here, we have used the LEP constraint: $v_A^2+v_B^2\gg (246 \, {\rm GeV})^2$ \cite{Carena:2004xs}. 
Note that only the two RHNs are involved in the seesaw mechanism (the so-called ``minimal seesaw'' \cite{MinSeesaw}), 
  while the third RHN ($N_R^3$) has no direct coupling with the SM fields and 
  hence it is naturally a dark matter (DM) candidate. 
Recently, this RHN DM scenario has been proposed in Ref.~\cite{Okada:2018tgy}. 
%For another attempt with the alternative charge assignment, 
%  see, for example, Ref.~\cite{Ma:2014qra}, where the inverse seesaw mechanism is implemented. 

Before going to our analysis for the 250 GeV ILC, 
  we first need to understand the current status and the future prospect 
  of the $B-L$ models in terms of the LHC experiments.  
The ATLAS and the CMS collaborations have been searching for a $Z^\prime$ boson resonance 
  with a variety of final states at the LHC Run-2 with $\sqrt{s}=13$ TeV 
  and the most severe upper bound relevant to our $Z^\prime$ boson production cross section  
%The current LHC Run-2 data show no evidence for such a resonance state and 
%  the upper bound on the $Z^\prime$ boson production cross sections have been obtained. 
%The most severe constraint relevant to our $Z^\prime$ boson is 
  has been obtained from the resonance search with a dilepton ($e^+e^-$ or $\mu^+ \mu^-$) final state. 
The latest results by the ATLAS collaboration \cite{ATLAS:2017}  and the CMS collaboration \cite{CMS:2017} 
   with a 36/fb integrated luminosity are consistent with each other and set 
   the lower mass bound of around 4.5 TeV for the sequential SM $Z^\prime$ boson. 
For our analysis, we employ the ATLAS result \cite{ATLAS:2017}.

%In our analysis, we interpret the current LHC constraint 
%
%to the $Z^\prime$ boson of our U(1)$_{B-L}$ models  
%  to obtain an upper bound on the U(1)$_{B-L}$ gauge coupling ($g_{BL}$) 
%   as a function of $Z^\prime$ boson mass.  
%For our analysis, we employ the ATLAS result \cite{ATLAS:2017}. 

In the $B-L$ models, the differential cross section for the process, 
   $pp \to Z^\prime +X \to \ell^{+} \ell^{-} +X$, where $\ell^{+} \ell^{-}=e^+ e^-$ or $\mu^+ \mu^-$, 
   with respect to the dilepton invariant mass $M_{\ell \ell}$ is given by 
\begin{eqnarray}
 \frac{d \sigma}{d M_{\ell \ell}}
 &=&  \sum_{q, {\bar q}}
 \int^1_ \frac{M_{\ell \ell}^2}{E_{\rm LHC}^2} dx \, 
  \frac{2 M_{\ell \ell}}{x E_{\rm LHC}^2}  
 f_q(x, Q^2)  \,  f_{\bar q} \left( \frac{M_{\ell \ell}^2}{x E_{\rm LHC}^2}, Q^2
 \right)  \nonumber \\
&&  \times  {\hat \sigma} (q \bar{q} \to Z^\prime \to  \ell^+ \ell^-) ,
\label{CrossLHC}
\end{eqnarray}
where $Q$ is the factorization scale (we fix $Q= m_{Z^\prime}$, for simplicity),  
 $E_{\rm LHC}=13$ TeV is the LHC Run-2 energy, 
 $f_q$ ($f_{\bar{q}}$) is the parton distribution function for quark (anti-quark), 
  and the cross section for the colliding partons is given by  
\bea 
{\hat \sigma}(q \bar{q} \to Z^\prime \to  \ell^+ \ell^-) =
\frac{g_{BL}^4}{324 \pi} 
\frac{M_{\ell \ell}^2}{(M_{\ell \ell}^2-m_{Z^\prime}^2)^2 + m_{Z^\prime}^2 \Gamma_{Z^\prime}^2}.  
\label{CrossLHC2}
\eea
Here, the total decay width of the  $Z^\prime$ boson ($\Gamma_{Z'}$) is given by 
\bea
\Gamma_{Z'} &=& 
 \frac{g_{BL}^2}{24 \pi} m_{Z^\prime} 
 \left[ 13+ \sum_{j=1}^3 Q_{N^j}^2 \left( 1-\frac{4 m_{N^j}^2}{m_{Z^\prime}^2} \right)^{\frac{3}{2}} 
 %\theta \left( \frac{m_{Z^\prime}^2}{m_{N^j}^2} - 4 \right)  
 \right], 
\label{width}
\eea
where we have neglected all SM fermion masses, 
  % $\theta$ is the Heaviside step function, 
  and $Q_{N^j}$ is the U(1)$_{B-L}$ charge of the RHN $N_R^j$. 
%For the minimal $B-L$ model, $Q_{N^j}=-1$, while $Q_{N^{1,2}}=-4$ and $Q_{N^3}=+5$ 
%  for the alternative $B-L$ model. 
%Our LHC analysis for the two $B-L$ models are essentially the same except for 
%  the difference between the decay width formulas.   
For the minimal (alternative) $B-L$ model, let us consider two benchmark (degenerate) mass spectra for the RHNs:  
  $m_{N^{1,2,3}} (m_{N^{1,2}})= m_N=50$ GeV and $100$ GeV.  
It has been recently shown in Ref.~\cite{Okada:2018tgy} that in the alternative $B-L$ model,  
   $N_R^3$ plays the role of DM in the Universe, reproducing the observed DM relic abundance 
   with $m_{N^{3}} \simeq m_{Z^\prime}/2$. 
Motivated by the discussion, we set $m_{N^{3}} \simeq m_{Z^\prime}/2$, 
   so that the $N^3$ contribution to $\Gamma_{Z^\prime}$ is neglected.

In our LHC analysis, we employ CTEQ6L~\cite{CTEQ} for the parton distribution functions 
  and calculate the cross section of the dilepton production through the $Z^\prime$ boson exchange in the $s$-channel.  
Neglecting the mass for the RHNs in our LHC analysis, the resultant cross section  
   is controlled by only two parameters: $g_{BL}$ and $m_{Z^\prime}$. 
To derive a constraint for these parameters from the ATLAS 2017 results \cite{ATLAS:2017}, 
   we follow the strategy in Refs.~\cite{OO1, SO}: 
   we first calculate the cross section of the process, $pp \to Z^\prime +X \to \ell^{+} \ell^{-} +X$, 
   for the sequential SM $Z^\prime$ boson and find a $k$-factor ($k = 1.31$)    
   by which our cross section coincides with the cross section for the sequential SM $Z^\prime$ boson
   presented in the ATLAS paper \cite{ATLAS:2017}. 
We employ this $k$-factor for all of our LHC analysis, and find an upper bound on $g_{BL}$ 
   as a function of $m_{Z^\prime}$ from the ATLAS 2017 results. 
For the prospect of the future constraints to be obtained after the HL-LHC experiment with the 3000/fb integrated luminosity, 
    we refer the simulation result presented in the ATLAS Technical Design Report  \cite{ATL-TDR}. 
Figure 4.20 (b) in this report shows the prospective upper bound on the cross section, 
   $pp \to Z^\prime +X \to e^{+} e^{-} +X$, as low as $10^{-5}$ fb over the range of $2.5 \leq m_{Z^\prime}\, [{\rm TeV}] \leq 7.5$, 
   which results in a lower bound on $m_{Z^\prime} > 6.4$ TeV for the sequential SM $Z^\prime$ boson.

%% %%%%%%%%%%%%%%%%%%%%%%%%%%%%%%%%
\begin{figure}[t]
\begin{center}
\includegraphics[scale=0.8]{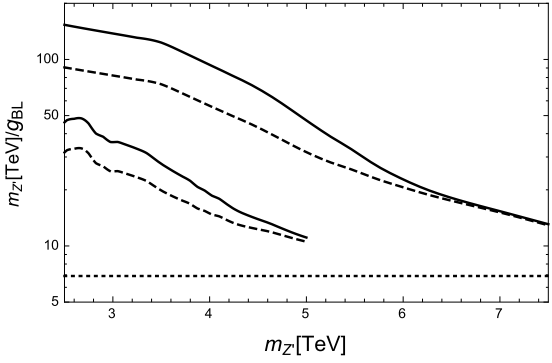}       
\end{center}
\caption{
The lower bounds on $m_{Z^\prime}/g_{BL}$ as a function of $m_{Z^\prime}$ 
   from the ATLAS 2017 result and the HL-LHC search reach \cite{ATL-TDR}, 
   along with the LEP constraint of $m_{Z^\prime}/g_{BL} > 6.9$ TeV (dotted horizontal line) \cite{Carena:2004xs}. 
}
\label{fig:LHC}
\end{figure}
%%%%%%%%%%%%%%%%%%%%%%%%%%%%%%%%%%

For the following ILC analysis, instead of the LHC upper bound on $g_{BL}$ as a function of $m_{Z^\prime}$, 
  it is more useful to plot the LHC lower bound on $m_{Z^\prime}/g_{BL}$, which is shown in Fig.~\ref{fig:LHC}. 
The lower and upper solid lines correspond to the lower bound from the ATLAS 2017 and 
  the prospective HL-LHC bound, respectively, for the minimal $B-L$ model. 
The corresponding lower  bounds for the alternative $B-L$ model are depicted as the dashed lines. 
In the alternative $B-L$ model, the $Z^\prime$ boson decay to a pair of RHNs dominates  
  the total decay width and hence the branching ratio into dileptons is relatively suppressed, 
  resulting in the LHC constraints weaker than those for the minimal $B-L$ model. 
Note that the LHC constraint for $m_{Z^\prime}/g_{BL}$ becomes dramatically weaker 
  as $m_{Z^\prime}$ increases.   
Since the ILC energy is much smaller than $m_{Z^\prime}$, 
  the $Z^\prime$ boson mediated processes at the ILC are described 
  by effective higher dimensional operators which are proportional to $(m_{Z^\prime}/g_{BL})^2$.  %\footnote{
%The effective 4-Fermi interactions mediated by a $Z^\prime$ boson have been searched 
%  by the LEP experiments \cite{LEPdata}. 
%No significant deviation from the SM predictions has been observed, and a lower bound on 
%   ${Z^\prime}/g_{BL}  > 7$ TeV has been obtained \cite{Carena:2004xs}.  
%We can see from the left panel of Fig.~\ref{fig:LHC} that the current LHC bounds are more stringent  
%   than the LEP bound.     
%}   
Therefore, the plots in Fig.~\ref{fig:LHC} imply that the ILC can be a more powerful machine 
  than the LHC to explore the $B-L$ models, 
  if the $Z^\prime$ boson mass is beyond the search reach of the HL-LHC experiment.

\begin{figure}[t]
\begin{center}
\includegraphics[scale=0.8]{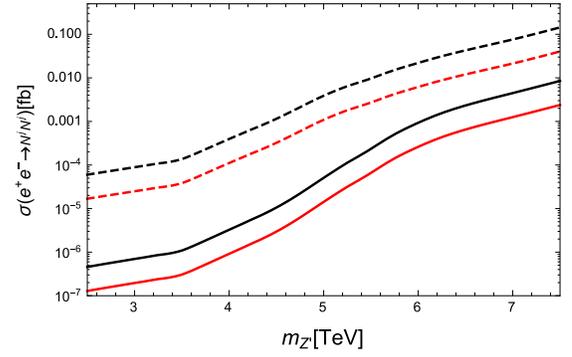}
\end{center}
\caption{
The RHN pair production cross sections at the 250 GeV ILC,  
  along the prospective HL-LHC bounds shown in Fig.~\ref{fig:LHC}. 
The upper (black) and lower (red) solid lines are the results for the minimal $B-L$ model 
  with $m_{N^{1,2,3}}=50$ GeV and $100$ GeV, respectively. 
Th results for the alternative $B-L$ model are shown as the upper (black) and lower (red) dashed lines 
  corresponding to $m_{N^{1,2}}=50$ GeV and $100$ GeV, respectively. 
}
\label{fig:ILC}
\end{figure}
%%%%%%%%%%%%%%%%%%%%%%%%%%%%%%%%%%

Let us now investigate the RHN pair production at the 250 GeV ILC. 
The relevant process is $e^+ e^- \to Z^{\prime *}\to N^i N^i$ mediated by a virtual $Z^\prime$ boson 
  in the $s$-channel. 
Since the collider energy $\sqrt{s}=250$ GeV is much smaller than $m_{Z^\prime}$,  
  the RHN pair production cross section is approximately given by 
\bea
&&\sigma(e^+ e^- \to Z^{\prime *}\to N^i N^i)  
\nonumber  \\
%  \sigma  =  
%   \frac{g_{BL}^4}{24 \pi}  (Q_{N^i})^2 \frac{s}{(s-m_{Z^\prime}^2)^2} \left( 1-\frac{4 m_{N^i}^2}{m_{Z^\prime}^2} \right)^{\frac{3}{2}}, 
%
%
&&\simeq \frac{(Q_{N^i})^2}{24 \pi} s  \left( \frac{g_{BL}}{m_{Z^\prime}}\right)^4
 \left( 1-\frac{4 m_{N^i}^2}{m_{Z^\prime}^2} \right)^{\frac{3}{2}}. 
%
%
%  \theta \left( \frac{m_{Z^\prime}^2}{m_{N^j}^2} - 4 \right)
% \nonumber \\ 
%
%&\simeq & 
%    \frac{ (Q_{N^j})^2}{24 \pi} \, s  \, \left(\frac{g_{BL}}{m_{Z^\prime}} \right)^4
%     \left( 1-\frac{4 m_{N^j}^2}{m_{Z^\prime}^2} \right)^{\frac{3}{2}} 
%  \theta \left( \frac{m_{Z^\prime}^2}{m_{N^j}^2} - 4 \right) ,
\label{ILC_Xsec}   
\eea
%where $\sqrt{s}=250$ GeV $\ll m_{Z^\prime}$ for the 250 GeV ILC. 
% and we have neglected the $Z^\prime$ decay width because of $s \ll m_{Z^\prime}^2$. 
%For fixed $m_{Z^\prime}$ and $m_{N^j}$ values, we have a maximum cross section at the ILC 
%   so as to satisfy the LHC constraints.  
%Note that the maximum cross section is increasing 
%   as the LHC lower bound on $m_{Z^\prime}/g_{BL}$ becomes weaker.   
%For a common mass choice of $m_{N^j}$ ($j=1,2$) for the minimal and the alternative $B-L$ models, 
%   the production cross section in the alternative case is $16$ times bigger than the minimal model 
%   because the cross section is proportional to $(Q_{N^j})^2$.   
%
For our benchmark RHN mass spectra, 
   we show in Fig.~\ref{fig:ILC} the RHN pair production cross sections 
   at the 250 GeV ILC, 
   along the prospective HL-LHC bounds on $m_{Z^\prime}/g_{BL}$ shown in Fig.~\ref{fig:LHC}. 
%For the minimal $B-L$ model, the maximum ILC cross sections for $m_{N^{1,2,3}}=50$ GeV and $100$ GeV
%    are depicted by the upper and lower solid lines, respectively, 
%  along the solid lines for the prospective HL-LHC bounds in Fig.~\ref{fig:LHC}. 
%The corresponding maximum cross sections for the alternative $B-L$ model 
%   are shown as the dashed lines (the upper dashed line corresponds to $m_{N^{1,2}}=50$ GeV). 
For $m_{Z^\prime}=7.5$ TeV, we have found 
  $\sigma(e^+ e^- \to Z^{\prime *} \to N^i N^i) =0.0085$  and $0.14$ fb 
  for $m_{N^{1,2,3}}=50$ GeV and $m_{N^{1,2}}=50$ GeV, 
  respectively, for the minimal and alternative $B-L$ models. 
For the degenerate RHN mass spectra, 
  we have $\sum_{i=1}^3\sigma(e^+ e^- \to Z^{\prime *}\to N^{i} N^{i}) =0.026$ fb and 
  $\sum_{i=1}^2\sigma(e^+ e^- \to Z^{\prime *} \to N^{i} N^{i}) =0.29$ fb for each model, 
  and thus 52 and 576 events with the 2000/fb goal luminosity of the 250 GeV ILC, 
  while satisfying the prospective constraints after the HL-LHC with the 3000/fb integrated luminosity. 
Considering the smoking-gun signature of the RHN pair production for which the SM backgrounds are few, 
  the 250 GeV ILC can operate as a Majorana RHN discovery machine 
  towards confirming the type-I seesaw mechanism. 
In the second stage of the ILC with $\sqrt{s}=500$ GeV \cite{Fujii:2017vwa} 
   we expect roughly 4 times more events with the same goal luminosity.

For detailed discussion about the ILC phenomenology, 
   we need to consider the decay processes of the heavy neutrinos. 
Assuming $|m_D^{ij} / m_{N^j}| \ll 1$ in Eq.~(\ref{mass_minimal}) or Eq.~(\ref{mass_alternative}), 
  the type-I seesaw mechanism leads to the light Majorana neutrino mass matrix of the form:  
\bea
m_{\nu} \simeq m_{D} M_{N}^{-1}m_{D}^{T} = \frac{1}{m_N} \, m_{D} \,m_{D}^{T}, 
\label{seesawI}
\eea 
where $M_N = m_N {\bf 1}$ with the $3\times3$ ($2 \times 2$) identity matrix ${\bf 1}$ 
  for the minimal (alternative) $B-L$ model. 
Through the seesaw mechanism, the SM neutrinos and the RHNs are mixed in the mass eigenstates. 
The flavor eigenstates of the SM neutrinos $(\nu)$ are expressed in terms of the light $(\nu_m)$ and heavy $(N_m)$ 
  Majorana neutrino mass eigenstates as 
$\nu \simeq  \mathcal{R} N_m + \mathcal{N} \nu_m $, 
  where $\mathcal{R} =m_D (M_N)^{-1}$, 
  $\mathcal{N}=\Big(1-\frac{1}{2}\mathcal{R}^\ast\mathcal{R}^T\Big)U_{\rm{MNS}}\simeq U_{\rm{MNS}}$, 
  and $U_{\rm{MNS}}$ is the neutrino mixing matrix  which diagonalizes the light neutrino mass matrix as
\bea
U_{\rm{MNS}}^T m_\nu U_{\rm{MNS}} ={\rm diag}(m_1, m_2, m_3).
\eea 
Through the mixing matrix $\mathcal{R}$ and the original Dirac Yukawa interactions, 
  the heavy neutrino mass eigenstates, if kinematically allowed, decay into $\ell W$, $\nu Z$, $\nu h$ ($h$ is the SM Higgs boson). 
%Here, we assume the $B-L$ Higgs bosons are all heavy with negligibly small mixings with the SM Higgs bosons. 
If the decays to on-shell $W$/$Z$/$h$ are not allowed, 
  the heavy neutrinos decay into SM fermions mainly through off-shell $W$/$Z$.  
%Since the couplings of the SM Higgs boson to the SM fermions are very small, 
%  we ignore the decay process through the off-shell Higgs boson. 
In Appendix I-III, we list the heavy neutrino decay width formulas 
  for two cases: (A) the heavy neutrinos decay into three SM fermions through off-shell $W$/$Z$, 
  and (B) the heavy neutrinos decay into $\ell W$, $\nu Z$, $\nu h$. 
As shown in Appendix IV, in our simple parametrization of $m_D$ from the type-I seesaw formula, 
   $|\mathcal{R}_{\alpha i}|^2$ is expressed as a function of only the lightest light neutrino mass eigenvalue $m_{\rm lightest}$ 
   and $m_N$ by using the neutrino oscillation data. 
Therefore, once we fix $m_{\rm lightest}$ and $m_N$, the heavy neutrino decay processes are completely determined.

%%%%%%%%%%%%%%%%%%%%%%%%%%%%%%%%%%%%%%%%%%%%%
\begin{table}[t]
\begin{center}
%\begin{tabular*}{0.4\textwidth}{@{\extracolsep{\fill}} c  |  c  |  c  |  c  |}
\begin{tabular}{| c | c |  c  |  c  |}
\hline
 $m_N=50$ GeV & $~e + jj~$ & $~\mu + jj~ $ & $~\tau + jj~$   \\ 
	\hline
  	$N^1$ & $0.412$ & $0.104$ & $0.104$  \\ 
	\hline
  	$N^2$ & $0.204$ & $0.224$ & $0.224$  \\ 
	\hline
  	$N^3$ & $0.0154 $ & $0.310$ & $0.310$  \\ 
	\hline
	\hline
 $m_N=100$ GeV & $e + jj$ & $\mu + jj $ & $\tau + jj$   \\ 
	\hline
  	$N^1$ & $0.587$ & $0.148$ & $0.148$  \\ 
	\hline
  	$N^2$ & $0.276$ & $0.304$ & $0.304$  \\ 
	\hline
  	$N^3$ & $0.0208$ & $0.431$ & $0.431$  \\ 
	\hline
\end{tabular}
\end{center}
\caption{
Branching ratios of the decay of the heavy neutrinos $N^{i=1,2,3}$ into $e/\mu/\tau + jj$ 
   in the minimal $B-L$ model.  
The resultant branching ratios are independent of the pattern of the light neutrino spectra 
   and $m_{\rm lightest}$.  
}
\label{tab:minimal}
\end{table}
%%%%%%%%%%%%%%%%%%%%%%%%%%%%%%%%%%%%%%%%%%%%%

We now consider the smoking-gun signature of the heavy neutrino pair production, namely, 
   $ e^+ e^- \to Z^{\prime *} \to N^i N^i$, 
   followed by  $N^i N^i \to \ell^{\pm} \ell^{\pm} W^{\mp (*)} W^{\mp (*)} \to \ell^{\pm} \ell^{\pm} jjjj$. 
This lepton number violating process originates from the Majorana nature of the heavy neutrinos 
   and is basically free from the SM background. 
The final same-sign dileptons can also violate the lepton flavor because of the neutrino mixing matrix.   
Using the formulas given in Appendix II-IV,  % and the matrix ${\cal R}$, 
   we calculate the branching ratios of the process, $N^i \to e/\mu/\tau + jj$. 
For the minimal $B-L$ model, the resultant branching ratios into  $N^i \to  \ell W^{(*)} \to \ell jj$ for each flavor charged lepton 
    are listed in Table \ref{tab:minimal}, for $m_N=50$ GeV and $100$ GeV.   
For the degenerate RHN masses, we find that the resultant branching ratios are independent 
   of the pattern of the light neutrino mass spectra and $m_{\rm lightest}$.  
We find the branching ratio of $N^i N^i \to \ell^{\pm} \ell^{\pm} jjjj$ for any lepton flavors to be about 20\%. 
For the alternative $B-L$ model, we obtain a similar result. See Appendix V for details.

Finally, let us discuss another interesting signature of the heavy neutrino production. 
Eq.~(\ref{seesawI}) indicates elements of ${\cal R}$ is very small, so that heavy neutrinos can be long-lived.  
Such long-lived heavy neutrinos leave displaced vertex signatures which can be easily distinguished 
   from the SM background events. 
For the minimal $B-L$ model, we show the decay lengths (lifetime times speed of light) of heavy neutrinos 
   in Appendix VI (see Figs.~\ref{fig:ctau50} and \ref{fig:ctau100}). 
Interestingly, the longest-lived heavy neutrino lifetime is inversely proportional to $m_{\rm lightest}$ \cite{Jana:2018rdf}, 
   so that $m_{\rm lightest}$ can be determined once the long-lived heavy neutrino is observed 
   with a displaced vertex. 
Note that this heavy neutrino becomes stable 
   and thus a DM candidate in the limit of $m_{\rm lightest} \to 0$. 
We can see that in this limit, a $Z_2$ symmetry comes out as an enhanced symmetry,  
   under which the DM particle is odd.   
Thus, the stability of the DM particle is ensured by this $Z_2$ symmetry, 
   as previously discussed in Ref.~\cite{OS}.

In conclusion, we have considered the minimal and the alternative $B-L$ models 
   which are simple and well-motivated extension of the SM to incorporate 
   the SM neutrino masses and flavor mixings through the type-I seesaw mechanism. 
Towards the experimental confirmation of the seesaw mechanism, 
   we have investigated the heavy neutrino pair production mediated 
   by the $Z^\prime$ boson at the 250 GeV ILC. 
The $Z^\prime$ boson mediated process is very severely constrained 
   by the LHC Run-2 results and the constraints will be more stringent in the future. 
Nevertheless, we have found that if $Z^\prime$ boson is very heavy, for example, $m_{Z^\prime} \gtrsim 7.5$ TeV, 
   the heavy neutrino pair production cross section at the 250 ILC can be sizable, 
   while satisfying the prospective bounds after the HL-LHC experiment 
   with the 3000/fb integrated luminosity. 
Once a pair of heavy neutrinos is produced, the same-sign dilepton final states can be observed, 
   which are the signature of the Majorana nature of the heavy neutrinos. 
In addition, the heavy neutrinos can be long-lived and leave displaced vertex signatures.  
Therefore, it is possible that the 250 GeV ILC operates as not only a Higgs Factory 
   but also a heavy neutrino discovery machine to explore the origin of the Majorana neutrino 
   mass generation, namely the seesaw mechanism.

The $Z^\prime$ boson can be indirectly searched with the dilepton final states, 
$e^+ e^- \to \ell^+ \ell^-$, at the 250 GeV ILC by observing a deviation of the total
cross section from its SM prediction. 
For $m_{Z^\prime}=7.5$ TeV,  we have obtained a deviation of ${\cal O}$(1 \%)  
from the SM prediction through an interference between the SM process 
and the $Z^\prime$ boson mediated process. 
This deviation can be explored at the ILC \cite{Fujii:2017vwa}.    

%%%%%%%%%%%%%%%%%%%%%%%%%%%%%%%%%%
\section*{Acknowledgments}
%%%%%%%%%%%%%%%%%%%%%%%%%%%%%%%%%%
This work is supported in part by 
the Japan Society for the Promotion of Science Postdoctoral Fellowship for Research in Japan (A.D.),  
the United States Department of Energy Grant (DE-SC0013680 (N.O.) and DE-SC0013880 (D.R.)), 
and the M. Hildred Blewett Fellowship of the American Physical Society, www.aps.org (S.O.).

%%%%%%%%%%%%%%%%%%%%%%%%%%%%%%%%%%
\section{Appendix}
\label{app}
%%%%%%%%%%%%%%%%%%%%%%%%%%%%%%%%%%
\subsection{I. Weak interactions of the neutrino mass eigenstates} 
%%%%%%%%%%%%%%%%%%%%%%%%%%%%%%%%%%
In terms of the neutrino mass eigenstates, the charged current $(\rm{CC})$ interaction can be written as 
\bea 
\mathcal{L}_{\rm{CC}}= 
 -\frac{g}{\sqrt{2}} W_{\mu}
  \overline{\ell_\alpha} \gamma^{\mu} P_L 
   \left( {\cal N}_{\alpha j} \nu_{m_j}+ {\cal R}_{\alpha j} N_{m_j} \right) + \rm{h.c.}, 
\label{CC}
\eea
where $\ell_\alpha$ ($\alpha=e, \mu, \tau$) denotes the three generations of the charged leptons, 
  and $P_L =  \frac{1}{2}(1- \gamma_5)$ is the left-handed projection operator.  
Similarly, the neutral current $(\rm{NC})$ interaction is given by 
\bea 
\mathcal{L}_{\rm{NC}}&=& 
 -\frac{g}{2 \cos \theta_{\rm W}}  Z_{\mu} 
\Big[ 
  \overline{\nu_{m_i}} \gamma^{\mu} P_L ({\cal N}^\dagger {\cal N})_{ij} \nu_{m_{j}} \nonumber \\  
&& +  \overline{N_{m_{i}}} \gamma^{\mu} P_L ({\cal R}^\dagger {\cal R})_{ij} N_{m_{j}} \nonumber \\
&& + \Big\{ 
  \overline{\nu_{m_{i}}} \gamma^{\mu} P_L ({\cal N}^\dagger  {\cal R})_{ij} N_{m_{j}} 
  + \rm{h.c.} \Big\} 
\Big] , 
\label{NC}
\eea
where $\theta_{\rm W}$ is the weak mixing angle.

%%%%%%%%%%%%%%%%%%%%%%%%%%%%%%%%%%
\subsection{II. Heavy neutrino decay to three SM fermions} 
%%%%%%%%%%%%%%%%%%%%%%%%%%%%%%%%%%
The heavy neutrinos are lighter than the weak bosons, they decay into three SM fermions 
   via off-shell $W$ and $Z$ bosons mediated processes. 
The partial decay widths into three lepton final states are as follows: 
\bea
\Gamma^{(W^*)}({N^i \to \ell_L^\alpha \ell_L^\beta \nu^\kappa})
 &=&  |R_{\alpha i}|^{2} \,
%\left(\sum_{\beta ,\kappa}
|U^{\beta \kappa }_{\rm MNS}|^2  
%\right) 
\,   \Gamma_{\rm N^i},
\nonumber \\
\Gamma^{(Z^*)}({N^i \to \nu^\alpha \ell_L^{\beta} \ell_L^{\kappa}})
&=&  |{\cal R}_{\alpha i}|^{2} \,
%\left(\sum_{\beta ,\kappa=1} 
\delta_{\beta \kappa} %\right) 
\cos^22\theta_W 
 \, \frac{1}{4} \, \Gamma_{\rm N^i},
\nonumber \\
\Gamma^{(Z^*)}({N^i \to \nu^\alpha \ell_R^{\beta} \ell_R^{\kappa}})
&=&  |{\cal R}_{\alpha i}|^{2} 
%\left(\sum_{\beta ,\kappa} 
\delta_{\beta \kappa} %\right) \,
  \sin^4\theta_W  \, \Gamma_{\rm N^i}, 
\nonumber \\
\Gamma^{(Z^*)}({N^i \to \nu^\alpha \nu^\beta \nu^\kappa})
 &=&  |{\cal R}_{\alpha i}|^{2} \, 
%\left(\sum_{\beta ,\kappa}
\delta_{\beta \kappa} %\right)  
 \, \frac{1}{4} \, \Gamma_{\rm N^i},
\label{eq:dwlepton1}
\eea
where 
\bea
\Gamma_{\rm N^i} =\frac{G_F^2}{192 \pi^3} m_{N^i}^5
\eea
with the Fermi constant $G_F$, and 
  $U_{\rm MNS}^{\beta \kappa}$ is a $(\beta, \kappa)$-element of the neutrino mixing matrix. 
%  and $\sum_{\beta ,\kappa}|U^{\beta \kappa }_{MNS}|^2 = 3 = \sum_{\beta ,\kappa} \delta_{\beta \kappa}$. 
In deriving the above formulas, we have neglected all lepton masses. 
For the lepton final states, we have an interference between the $Z$ and $W$ boson mediated decay processes: 
\bea
\Gamma^{(Z^*/W^*)}({N^i \to \nu^\alpha \ell^\alpha \ell^\alpha})
 =|{\cal R}_{\alpha i}|^{2} \, 
 2{\rm Re}[U_{\rm MNS}^{i i}] \, \Gamma_{\rm N^i}.
\label{eq:dwlepton2}
\eea

The partial decay widths into one lepton plus two quarks are as follows: 
\bea
\Gamma^{(W^*)}(N^i \to \ell^\alpha  q_L^\beta{\bar q}_L^\kappa)
 &=&N_c \times |{\cal R}_{\alpha i}|^{2}  \,
%\left(\sum_{\beta, \kappa}
|V^{\beta \kappa}_{\rm CKM}|^2 
%\right)
\, \Gamma_{\rm N^i},
\nonumber \\
\Gamma^{(Z^*)}(N^i \to \nu^\alpha  q_L^\beta{\bar q}_L^\kappa)
 &=&  N_c \times|{\cal R}_{\alpha i}|^{2} \,   
%\left(\sum_{\beta ,\kappa}
\delta_{\beta \kappa}  %\right) 
\, Q_L^2\, \Gamma_{\rm N^i},
\nonumber \\
\Gamma^{(Z^*)}(N^i \to \nu^\alpha  q_R^\beta{\bar q}_R^\kappa)
 &=&N_c \times |{\cal R}_{\alpha i}|^{2}  \, 
%\left(\sum_{\beta ,\kappa} 
\delta_{\beta \kappa}    %\right) 
\, Q_R^2 \, \Gamma_{\rm N^i},
\label{eq:dwquark}
\eea
where $N_c = 3$ is the color factor, $V_{\rm CKM}^{\beta \kappa}$ is a $(\beta, \kappa)$-element of the quark mixing matrix, 
   $Q_{L} =  1/2 - (2/3) \sin^2\theta_W$ and  $Q_{R} =  - (2/3) \sin^2\theta_W$ for a up-type quark, 
   and $Q_{L} =  -1/2 - (1/3) \sin^2\theta_W$ and  $Q_{R} =  - (1/3) \sin^2\theta_W$ for a down-type quark. 
%There is no interference between $W$ and $Z$ boson mediated processes. 
%Also, we fix $m_N \leq 40$ GeV, hence we only consider the first two generation 
%of quarks in the final states and $\sum_{\beta, \kappa} |V^{q^\beta {\bar q}^\kappa}_{CKM}|^2 = 2$. 

%%%%%%%%%%%%%%%%%%%%%%%%%%%%%%%%%%
\subsection{III. Heavy neutrino decay to on-shell $W$/$Z$/$h$} 
%%%%%%%%%%%%%%%%%%%%%%%%%%%%%%%%%%
If the heavy neutrinos are heavy enough to decay into $\ell W$, $\nu_{\ell} Z$, and $\nu_{\ell} h$,  
  the partial decay widths are as follows: 
\bea
\Gamma(N_m^i \rightarrow \ell_{\alpha} W)
 &=& \frac{1}{16 \pi} 
 \frac{ (M_{N}^2 - m_W^2)^2 (M_{N}^2+2 m_W^2)}{M_{N}^3 v_h^2} \, |{\cal R}_{\alpha i}|^{2},
\nonumber \\
\Gamma(N_m^i \rightarrow \nu_{\ell_{\alpha}} Z)
 &=& \frac{1}{32 \pi} 
 \frac{ (M_{N}^2 - m_Z^2)^2 (M_{N}^2+2 m_Z^2)}{M_{N}^3 v_h^2} \, |{\cal R}_{\alpha i}|^{2},
\nonumber \\
\Gamma(N_m^i \rightarrow \nu_{\ell_{\alpha}} h)
 &=& \frac{1}{32 \pi}\frac{(M_{N}^2-m_h^2)^2}{M_{N} v_h^2} \, |{\cal R}_{\alpha i}|^{2}. 
\label{eq:dwofshell}
\eea

%%%%%%%%%%%%%%%%%%%%%%%%%%%%%%%%%%
\subsection{IV. Determining $R_{\alpha i}$ from the neutrino oscillation data} 
%%%%%%%%%%%%%%%%%%%%%%%%%%%%%%%%%%
The elements of the matrix ${\cal R}$ are constrained so as to reproduce the neutrino oscillation data. 
In our analysis, we adopt the following values for the neutrino oscillation parameters: 
   $\Delta m_{12}^2 = m_2^2-m_1^2 = 7.6 \times 10^{-5}$ eV$^2$, 
   $\Delta m_{23}^2= |m_3^2-m_2^2|=2.4 \times 10^{-3}$ eV$^2$, 
   $\sin^2 2\theta_{12}=0.87$, $\sin^2 2\theta_{23}=1.0$,
   and  $\sin^{2}2{\theta_{13}}=0.092$ \cite{PDG}. 
The neutrino mixing matrix is explicitly given by
\bea
&&U_{\rm{MNS}} = \nonumber \\
&& \begin{pmatrix} c_{12} c_{13}&c_{12}c_{13}&s_{13}e^{-i\delta}\\-s_{12}c_{23}-c_{12}s_{23}s_{13}e^{i\delta}&c_{12}c_{23}-s_{12}s_{23}s_{13}e^{i\delta}&s_{23} c_{13}\\ s_{12}c_{23}-c_{12}c_{23}s_{13}e^{i\delta}&-c_{12}s_{23}-s_{12}c_{23}s_{13}e^{i\delta}&c_{23}c_{13}
\end{pmatrix} \nonumber \\
&& \times 
\begin{pmatrix}
1&0&0\\
0&e^{-i \rho_1}&0\\
0&0&e^{-i \rho_2}
 \end{pmatrix},
 \nonumber 
\eea
where $c_{ij}=\cos\theta_{ij}$,  $s_{ij}=\sin\theta_{ij}$, and $\rho_1$ and $\rho_2$ 
  are the Majorana phases ($\rho_2=0$ in the minimal seesaw).  
For simplicity, we set the Dirac $CP$-phase as $\delta=3\pi/2$ from the indications 
 by the recent T2K \cite{Abe:2017uxa} and NO$\nu$A \cite{Adamson:2016tbq} data.

In our analysis we consider two patterns of the light neutrino mass spectrum, 
  namely the Normal Hierarchy (NH) where the light neutrino mass eigenvalues are ordered as
  $m_1 < m_2 < m_3$ and the Inverted Hierarchy (IH) 
  where the light neutrino mass eigenvalues are ordered as $m_3 < m_1< m_2$. 
In the NH (IH) case, this lightest mass eigenvalue $m_{\rm lightest} $ is identified with $m_1$ $(m_3)$. 
Thus, the mass eigenvalue matrix for the NH case is expressed as
\bea 
  D_{\rm{NH}} ={\rm diag}
  \left(m_{\rm lightest}, m_2^{\rm{NH}}, m_3^{\rm{NH}} \right),  
\label{DNH}
\eea 
with $m_2^{\rm{NH}}=\sqrt{ \Delta m_{12}^2+m_{\rm lightest}^2}$ 
        and $m_3^{\rm{NH}}=\sqrt{\Delta m_{23}^2 + (m_2^{\rm{NH}})^2}$, 
while the mass eigenvalue matrix for the IH case is 
\bea 
  D_{\rm{IH}} ={\rm diag}
\left( m_1^{\rm{IH}}, m_2^{\rm{IH}}, m_{\rm lightest} \right) 
\label{DIH}
\eea 
with $m_2^{\rm{IH}}=\sqrt{ \Delta m_{23}^2 + m_{\rm lightest}^2}$ 
     and $m_1^{\rm{IH}}=\sqrt{(m_2^{\rm{IH}})^2- \Delta m_{12}^2}$. 
%In our benchmarks for the ILC analysis, we consider the degenerate mass spectrum for 
%  the heavy neutrinos, so that we parametrize $M_N = m_N {\bf 1}$ 
%  with the $3\times3$ ($2 \times 2$) identity matrix ${\bf 1}$ for the minimal (alternative) $B-L$ model. 
Through the type-I seesaw mechanism,  the light neutrino mass matrix is expressed as 
\bea 
   m_\nu = m_D M_N^{-1}m_D^T 
   = U_{\rm{MNS}}^* D_{\rm{NH/IH}} U_{\rm{MNS}}^\dagger ,  
\eea  
for the NH/IH cases, respectively.
This formula allows us to simply parametrize the mixing matrix $\mathcal{R}$ as 
\bea
\mathcal{R}^{\rm{NH/IH}} = \frac{1}{\sqrt{m_N}} \, U_{\rm{MNS}}^{\ast} \, \sqrt{D_{\rm{NH/ IH}}},
\label{gp}
\eea
where $\sqrt{D_{\rm{NH}}} ={\rm diag}\left( \sqrt{m_{\rm lightest}}, \sqrt{m_2^{\rm{NH}}}, \sqrt{m_3^{\rm{NH}}} \right)$, and   
  $\sqrt{D_{\rm{IH}}} ={\rm diag}\left( \sqrt{m_1^{\rm{IH}}}, \sqrt{m_2^{\rm{IH}}}, \sqrt{m_{\rm lightest}} \right)$ 
  in the minimal $B-L$ model. 
For the minimal seesaw in the alternative $B-L$ model, 
   only two RHNs are involved in the seesaw mechanism and $m_{\rm lightest}=0$. 
In this case, $\sqrt{D_{\rm{NH/ IH}}}$ is expressed as $3 \times 2$ matrices as follows: 
\bea   
&&\sqrt{D_{\rm{NH}}} = 
\begin{pmatrix}
0&0\\
\sqrt{m_2^{\rm{NH}}} & 0\\
0& \sqrt{m_3^{\rm{NH}}}
 \end{pmatrix},  \nonumber \\
&& \sqrt{D_{\rm{IH}}} = 
\begin{pmatrix}
\sqrt{m_1^{\rm{IH}}} & 0\\
0& \sqrt{m_2^{\rm{IH}}} \\
0&0\\
 \end{pmatrix}.    
\label{3by2} 
\eea
With the inputs of the oscillation data, the mixing matrix ${\cal R}$ is found 
   to be a function $m_{\rm lightest}$, $m_N$ and the Majorana $CP$-phases. 
We find $|{\cal R}_{\alpha i}|^2$ is independent of the Majorana $CP$-phases, 
   so that the heavy neutrino decay processes are determined by only two free parameters:
   $m_{\rm lightest}$ and $m_N$.

%%%%%%%%%%%%%%%%%%%%%%%%%%%%%%%%%%%%%%%%%%%%%
\begin{table}[t]
\begin{center}
%\begin{tabular*}{0.4\textwidth}{@{\extracolsep{\fill}} c  |  c  |  c  |  c  |}
\begin{tabular}{| c | c |  c  |  c  |}
\hline
\multicolumn{4}{|c|}{NH case} \\
\hline 
\hline
 $m_N=50$ GeV & $~e + jj~$ & $~\mu + jj~ $ & $~\tau + jj~$   \\ 
	\hline
  	$N^1$ & $0.194$ & $0.213$ & $0.213$  \\ 
	\hline
  	$N^2$ & $0.0154$ & $0.318$ & $0.318$  \\ 
	\hline
	\hline
 $m_N=100$ GeV & $e + jj$ & $\mu + jj $ & $\tau + jj$   \\ 
	\hline
  	$N^1$ & $0.276$ & $0.304$ & $0.304$  \\ 
	\hline
  	$N^2$ & $0.0208$ & $0.431$ & $0.431$  \\ 
	\hline  
	\hline
	\hline	
\multicolumn{4}{|c|}{IH case} \\	
\hline
\hline 
$m_N=50$ GeV & $~e + jj~$ & $~\mu + jj~ $ & $~\tau + jj~$   \\ 
	\hline
  	$N^1$ & $0.412$ & $0.104$ & $0.104$  \\ 
	\hline
  	$N^2$ & $0.204$ & $0.224$ & $0.224$  \\ 
	\hline
	\hline
 $m_N=100$ GeV & $e + jj$ & $\mu + jj $ & $\tau + jj$   \\ 
	\hline
  	$N^1$ & $0.587$ & $0.148$ & $0.148$  \\ 
	\hline
  	$N^2$ & $0.276$ & $0.304$ & $0.304$  \\ 
	\hline
\end{tabular}
\end{center}
\caption{
Branching ratios of the heavy neutrinos $N^{i=1,2}$ into $e/\mu/\tau + jj$ 
   in the alternative $B-L$ model.  
}
\label{tab:alt}
\end{table}
%%%%%%%%%%%%%%%%%%%%%%%%%%%%%%%%%%%%%%%%%%%%%

%%%%%%%%%%%%%%%%%%%%%%%%%%%%%%%%%%%%%%%
\subsection{V. Heavy neutrino branching ratios in the alternative $B-L$ model} 
%%%%%%%%%%%%%%%%%%%%%%%%%%%%%%%%%%%%%%%
In the alternative $B-L$ model, only two RHNs are involved in the seesaw mechanism 
  and the mixing matrix ${\cal R}$ is given by Eq.~(\ref{gp}) with the $3 \times 2$ matrices in Eq.~(\ref{3by2}). 
It is easy to find a relation between ${\cal R}_{\alpha i}$ ($i=1,2,3$) in the minimal $B-L$ model 
  and ${\cal R}_{\alpha i}$ ($i=1,2$) in the alternative $B-L$ model (for vanishing Majorana phases). 
For the NH case, the element ${\cal R}_{\alpha i}$ in the alternative $B-L$ model 
  is the same as the element ${\cal R}_{\alpha i+1}$ in the minimal $B-L$ model. 
Similarly, for the IH case, the element ${\cal R}_{\alpha i}$ in the alternative $B-L$ model 
  is the same as the element ${\cal R}_{\alpha i}$ in the minimal $B-L$ model. 
For the alternative $B-L$ model the resultant branching ratios are listed in Table \ref{tab:alt}, 
  corresponding to Table \ref{tab:minimal} for the minimal $B-L$ model. 
Because of the relation between ${\cal R}$ elements in the two $B-L$ models,  
   the NH (IH) case results for $N^{1,2}$ in Table \ref{tab:alt} for $m_N=100$ GeV 
   are the same as those for $N^{2,3}$ ($N^{1,2}$) in Table \ref{tab:minimal}. 
This correspondence is not exact for the case of $m_N=50$ GeV, 
   since the partial decay width of Eq.~(\ref{eq:dwlepton2}) from the interference 
   contributes to the total decay width. 
We find that this contribution is small, and the correspondence is satisfied as a good approximation.      
Similarly to the minimal $B-L$ model, we find the branching ratio of $N^i N^i \to \ell^{\pm} \ell^{\pm} jjjj$ 
   for any lepton flavors to be about 20\%.

%%%%%%%%%%%%%%%%%%%%%%%%%%%%%%%%%%
\subsection{VI. Long-lived heavy neutrinos}  
%%%%%%%%%%%%%%%%%%%%%%%%%%%%%%%%%%

%% %%%%%%%%%%%%%%%%%%%%%%%%%%%%%%%%
\begin{figure}[h]
\begin{center}
\includegraphics[scale=0.8]{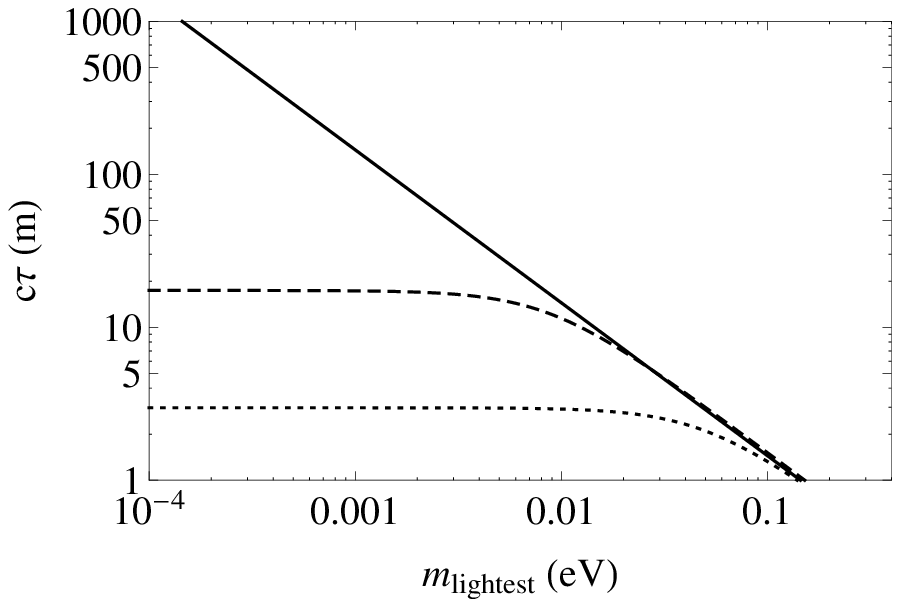}\\
\includegraphics[scale=0.8]{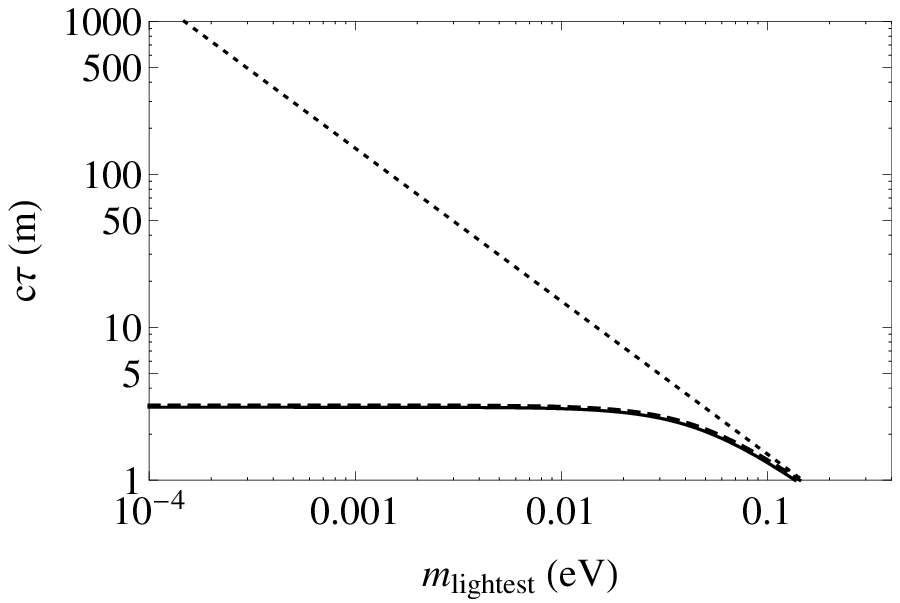}       
\end{center}
\caption{
Top panel: The lifetime (times speed of light) of $N^1$ (solid), $N^2$ (dashed) and $N^3$ (dotted) 
   for the NH light neutrino mass spectrum, for $m_N=50$ GeV.  
Bottom panel: Same as the top panel but for the IH light neutrino mass spectrum. 
}
\label{fig:ctau50}
\end{figure}
%%%%%%%%%%%%%%%%%%%%%%%%%%%%%%%%%%

%% %%%%%%%%%%%%%%%%%%%%%%%%%%%%%%%%
\begin{figure}[h]
\begin{center}
\includegraphics[scale=0.8]{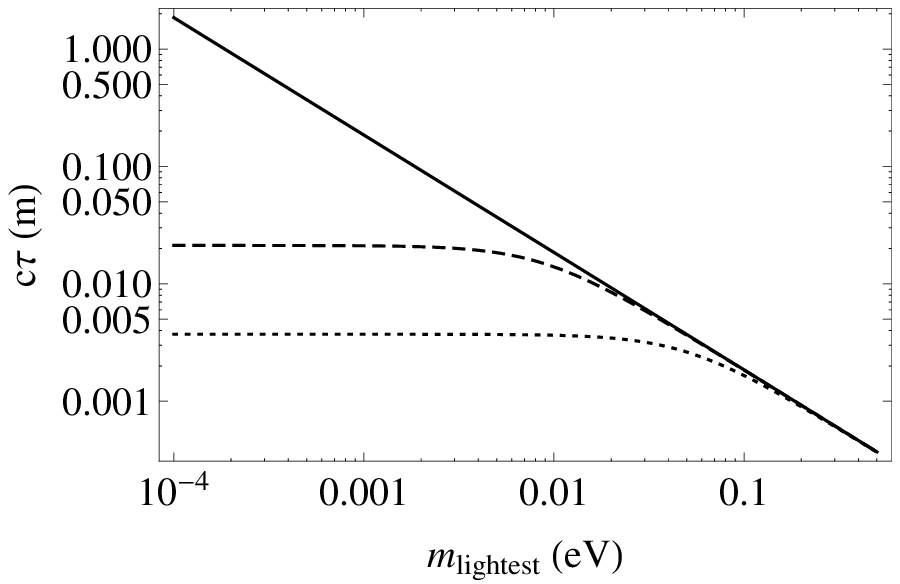}\\
\includegraphics[scale=0.8]{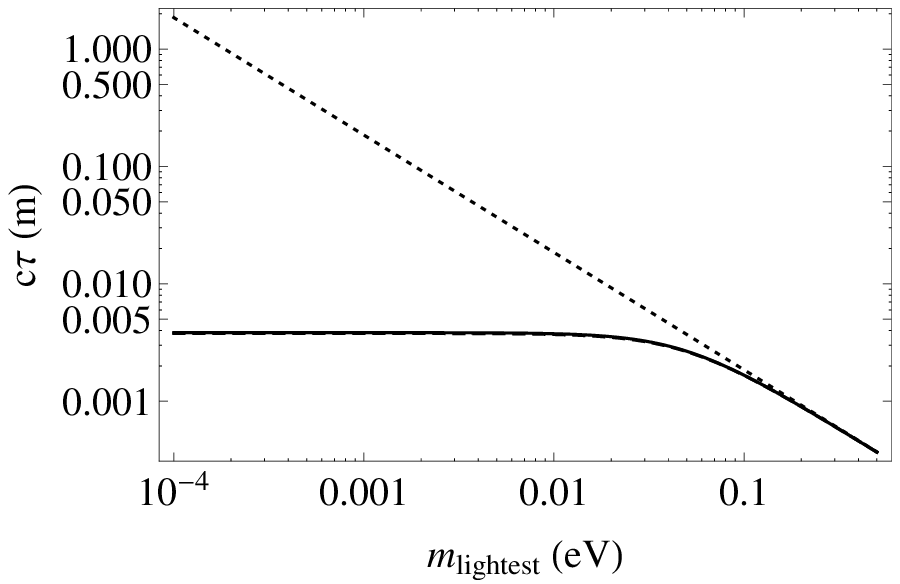}       
\end{center}
\caption{
Same as Fig.~\ref{fig:ctau50} but for $m_N=100$ GeV. 
}
\label{fig:ctau100}
\end{figure}
%%%%%%%%%%%%%%%%%%%%%%%%%%%%%%%%%%

In the minimal $B-L$ model, we calculate the total decay widths for $N^{1,2,3}$ 
   as a function of $m_{\rm lightest}$.  
We show in Fig.~\ref{fig:ctau50} the lifetime of $N^{1,2,3}$ for the NH (top) and IH (bottom) cases 
   for $m_N=50$ GeV.  
Fig.~\ref{fig:ctau100} is same as Fig.~\ref{fig:ctau50} but for $m_N=100$ GeV.    
The longest-lived heavy neutrino lifetime is inversely proportional to $m_{\rm lightest}$, 
   and hence it becomes a DM candidate in the limit of $m_{\rm lightest} \to 0$.

Similarly to our discussion about the branching ratios,  
   the lifetime of $N^{1,2}$ in the alternative $B-L$ model 
   can be obtained from the results in Figs.~\ref{fig:ctau50} and \ref{fig:ctau100}. 
The lifetime of $N^{1,2}$ for the NH case is given by the lifetime of $N^{2,3}$, respectively, 
   in the limit of $m_{\rm lightest} \to 0$. 
For the IH case, the lifetime of $N^{1,2}$ corresponds to the lifetime of $N^{1,2}$, respectively, 
   in the limit of $m_{\rm lightest} \to 0$. 
However, we have to be careful. 
These results are true only if $v_\nu=246$ GeV in Eq.~(\ref{mass_alternative}). 
In the alternative $B-L$ model, the neutrino Dirac mass is generated 
  from the VEV of the new Higgs doublet $H_\nu$ which only couples with neutrinos. 
This structure is nothing but the one in the so-called neutrinophilic two Higgs doublet model \cite{nuTHDM}. 
In order to avoid a significant change of the SM Yukawa couplings, 
  we normally take $v_\nu \ll v_h \simeq 246$ GeV. 
This means that the actual lifetime of $N^{1,2}$ is shorten by a factor of $(v_\nu/v_h)^2 \ll 1$. 
However, $N^1$ or $N^2$ can still be long-lived.

%%%%%%%%%%%%%%

%%%%%%%%%%%%%%%%%%%%%%%%%%%%%%%%%%%%%%

\end{document}